\documentclass[aps,prb,twocolumn,preprintnumbers,amsmath,amssymb]{revtex4}

\usepackage{graphicx}
\usepackage{dcolumn}
\usepackage{bm}
\usepackage{textcomp}

\begin{document}

\title{High-sensitivity microwave vector detection at extremely low-power levels for low-dimensional electron systems}

\author{W.\ H.\ Hsieh and C.\ H.\ Kuan}
\affiliation{Graduate Institute of Electronics Engineering and
Department of Electrical Engineering, National Taiwan University,
Taipei, Taiwan, Republic of China}
\author{Y.\ W.\ Suen,$^{\textrm{a)}}$ \footnotetext{$^{\textrm{a)}}$ Author to whom correspondence should be addressed; electronic mail:
ysuen@phys.nchu.edu.tw} S.\ Y.\ Chang, and L.\  C.\ Li}
\affiliation{Department of Physics, National Chung Hsing
University, 250, Kuo Kuang Road, Taichung 402, Taiwan, Republic of
China}
\author{B.\ C.\ Lee and C.\ P.\ Lee}
\affiliation{Department of Electronics Engineering, National Chiao
Tung University, Sinchu, Taiwan, Republic of China}

\date{\today}
\begin{abstract}
We present a high-sensitivity microwave vector detection system
for studying the low-dimensional electron system embedded in the
gaps of a coplanar waveguide at low temperatures. Using this
system, we have achieved 0.005\% and 0.001\textdegree\ resolutions
in amplitude and phase variations, respectively, at 10 GHz in a
magnetotransport measurement on  a quantum-wire array with an
average signal power less than $-$75 dBm into the sample at 0.3 K.
From the measured phase variation, we can distinguish a very tiny
change in the induced dipole moment of each quantum wire.

\end{abstract}
\maketitle

Coplanar waveguides (CPW's) have been successfully used as
broadband sensors in investigating the high-frequency
magnetotransport phenomena of low-dimensional electron systems
(LDES's), such as two-dimensional electron systems
(2DES's),\cite{Engel1993,Li1997,Ye2002,Lewis2002,Chen2003} and
anti quantum dots (QD's),\cite{Ye2002B} etc.  In these works, a
commercial vector network analyzer (VNA) is the major tool to
measure the variation of the propagation constant, including the
attenuation constant ($\alpha$) and the phase constant ($\beta$),
of the CPW that containing the active LDES in the gaps between the
metal electrodes. From $\alpha$ and $\beta$ one can extract the
longitudinal conductivity ($\sigma_{xx}$)\cite{Engel1993} (both
real and imaginary parts) of the LDES. However, since the
microwave power delivered to samples at temperature ($T$) below
few hundred mK must be very low, the resolution of the data
becomes very poor, especially for the phase part. Thus in most of
the previous studies using CPW sensors, they only presented
Re\{$\sigma_{xx}$\} data derived from $\alpha$ and discarded the
phase part. Even though Hohls \textit{et al.}\cite{Hohls2001}\ and
Lewis \textit{et al.}\cite{Lewis2001}\ have addressed the
Im\{$\sigma_{xx}$\} behavior in the integer quantum Hall (IQH)
regime based on other techniques, still, the resolution of
Im\{$\sigma_{xx}$\} is mediocre due to the constrain of VNA's.
Nevertheless, Im\{$\sigma_{xx}$\}, proportional to the change of
the real part of dielectric constant, gives important information
about the electric polarization, that is of special interest in
the case of QD's, quantum wires (QW's), or 2DES's at high magnetic
fields ($B$). Furthermore, the relatively small effective area of
QD's or QW's compared to 2DES samples leads to a very small signal
variation (or dynamic range), that makes the conventional VNA
measurement very difficult and impractical.

In this letter, we present a new detection scheme and the
instrumental implementation, which can resolve very small
variations not only in the amplitude but also the phase of an
extremely low-power-level microwave signal traveling through a CPW
with LDES's embedded in the gaps while some external sample
parameters, such as the applied magnetic field ($B$) or $T$, etc.,
is changed. The data of a low-$T$ magnetotransport measurement on
 a QW-array sample manifest the high-resolution
capability of this system.

A simplified schematic diagram to illustrate the principle of
 phase detection by a phase-lock loop\cite{PLL} (PLL) is depicted in Fig.\ 1 (a). The CPW sample is
connected to a PLL through two semirigid coaxial cables of total
length $L$. The PLL will force the total phase change ($\Delta
\phi$), including the phase change of the semirigid cables
($\Delta \phi_{L}$) and the CPW sample ($\Delta \phi_{s}$), to be
0 by tuning the frequency ($f$) of the voltage-controlled
oscillator (VCO) during the experiments, that is, $\Delta \phi =
\Delta \phi_{L}+\Delta \phi_{s}=0$, or $\Delta \phi_{s}=-\Delta
\phi_{L}$. Hence $\Delta \phi_{s}$ can be obtained directly from
the frequency change ($\Delta f$) of VCO via
\begin{eqnarray}
\Delta \phi_{s}=-\Delta \phi_{L}=-2\pi \Delta fL/v_{L}=-\Delta
\omega\tau_{L},
\end{eqnarray}
where $v_{L}$ is the phase velocity of the signal in the cable.
The result can be expressed as the product of the change of the
angular frequency ($\Delta \omega$) and the delay time
($\tau_{L}$) of the connecting cables with a different sign.

\begin{figure}
\includegraphics[width=7.5cm]{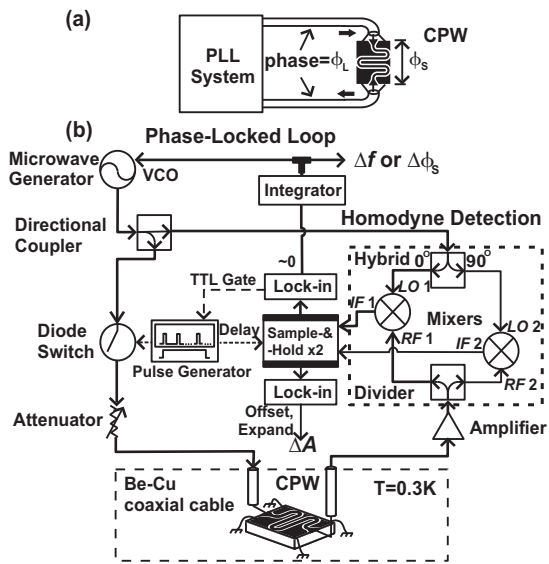}\\
\caption{(a) Simplified schematic diagram for  phase detection
using a PLL. (b) Block diagram of the vector detection system. The
meandering CPW containing a LDES in the gaps is part of the
microwave signal path in the PLL.}
\end{figure}

A complete block diagram, including the pulse handling circuits,
the microwave PLL, and the amplitude readout circuit, together
with the CPW sample in a cryogenic environment, is shown in Fig.\
1 (b). The microwave part of this system is basically a pair of
homodyne detectors (mixers) with reference signals of quadrature
phase difference. One of the mixers with 0\textdegree\ reference
(\textit{LO}1), used as the phase sensitive detector (PSD), has
zero output (\textit{IF}1) forced by the PLL, while the other one
with 90\textdegree\ reference (\textit{LO}2) has an output
(\textit{IF}2) proportional to the amplitude of the signal.

Besides homodyne detection we employ a double-pulse modulation
scheme to detect and average the microwave signal. A short pulse
train with a 0.2$\sim$2 $\mu$s pulse width and a 0.1$\sim$10\%
duty cycle, provided by a pulse generator and gated by a slow
square-wave TTL signal with a period of 1$\sim$10 ms from a
lock-in amplifier, modulates the microwave signal sent to the
sample. A time-delayed pulse with a 0.1$\sim$1 $\mu$s pulse width
triggered by the modulating pulses controls a sample-and-hold
(S\&H) circuit that samples the
 IF output of the microwave mixer. The holding
capacitor in the S\&H circuit is discharged through an analog
switch  when the TTL gating signal is low. Finally the lock-in
amplifier reads the output of the S\&H circuit. There are two sets
of pulse averaging circuits, one for the PLL and the other for the
amplitude readout. Note that the time constant of the lock-in
amplifier for the PLL is about 1$\sim$30 ms, in contrast to 300 ms
or 1 s for the amplitude readout part. The average of the PSD
output (\textit{IF}1) is sent to an integrator (loop filter of
PLL) with a time constant of 26.3 ms. The output of the integrator
connects to the FM input of the microwave source (VCO), thus
closing the PLL. In fact this PLL system is modified from what
people used in surface-acoustic-wave detection
experiments,\cite{Wixforth1989} but with improved pulse averaging
and amplitude detection methods.

We use three sets of microwave modules to cover the frequency from
about 60 MHz to 18 GHz. The details of our instrumentations and
circuit designs will be published elsewhere. To gain an idea of
the detection limit, we tested our system with only an 11 m
semirigid cable connected to the PLL without samples. The input
microwave signal is attenuated down to $-$70 dBm peak power, and
less than $-$90 dBm in average. The background phase fluctuation
we obtain in this test is less than 0.0003\textdegree\
(root-mean-squared value) for $f$$\lesssim$6 GHz and
0.006\textdegree\ for 6$\lesssim$$f$$\lesssim$18 GHz, which is
remarkably low for such a low-power signal. In fact, the signal
power reaching the low-noise amplifier (LNA) is even lower than
the input value claimed above due to the loss of the cable, which
is about $-$9 dB at 1 GHz and raises to $-$41 dB at 10 GHz. This
may explain why the noise in phase increases at high frequencies.
The resolution with a low-$T$ sample loaded is slightly worse due
to the loss of the sample and extra noise from the cryogenic
environment. The resolution of the amplitude readout for a
small-variation signal can be enhanced by the use of the "offset"
and "expand" functions of the lock-in amplifier.\cite{lockin}

In the following we will present measured results for a QW array
sample to demonstrate the resolving power of this method. The
sample is fabricated from a standard MBE-grown modulation-doped
GaAs/AlGaAs heterostructure containing a 2DES, which is 150 nm
under the surface.  The 2DES has a mobility of about
1.5$\times$$10^{5}$ cm$^{2}/$Vs at 4K, and a density of
1.1$\times$$10^{11}$ cm$^{-2}$. Before evaporating the
Cr/Au(10/300 nm) metal layers for the CPW pattern, we etch away
the 2DES part underneath. The widths of the center conductor and
the gap of the 50 $\Omega$ CPW are 36 $\mu$m and 23 $\mu$m,
respectively. A meandering pattern\cite{Engel1993} is used to
increase the effective length of the CPW. Subsequently we pattern
the 2DES left in the gap into about 7000 identical QW mesas, each
of 0.7 $\mu$m wide and 20 $\mu$m long, by using e-beam lithography
and chemical etching [Fig.\ 2 (a)]. The QW's, parallel to the
propagating direction of microwave signals, occupy only about 6 mm
in length of the straight sections of the meandering CPW.

The CPW sample is immersed in liquid $^3$He (0.3 K) with applied
$B$ perpendicular to the sample surface. The total time delay
given by the connection cables and microwave modules is 51.1 ns.
From Eq.\ 1, this time delay multiplied by $\Delta f$ gives the
phase change ($\Delta \phi_{s}$) of the CPW sample. Here $\Delta
f$ can be either obtained from the output voltage of the
integrator in the PLL scaled with the FM deviation setting of the
VCO, or measured directly with a microwave counter. The peak power
of the pulsed microwave signals is $-$50 dBm (about $-$66 dBm in
average) at the input end of the semirigid cable.

\begin{figure}
\includegraphics[width=7.5cm]{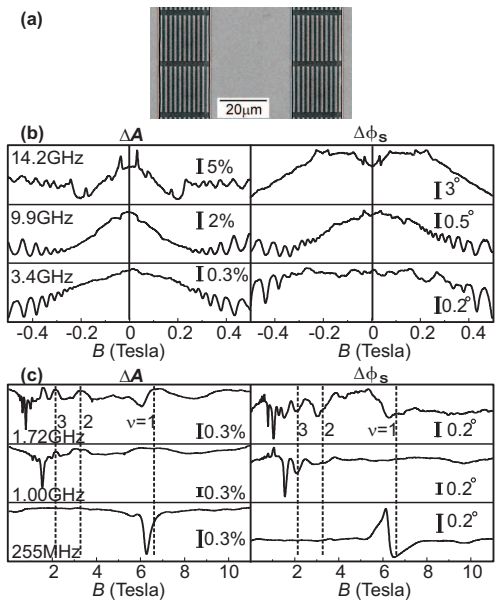}\\
\caption{(a) Simplified schematic diagram for  phase detection
using a PLL. (b) Block diagram of the vector detection system. The
meandering CPW containing a LDES in the gaps is part of the
microwave signal path in the PLL.}
\end{figure}

Figure 2 (b) displays low-field results for both directions of
$B$. Besides the apparent Shubinkov-de-Hass oscillations, we can
observe additional intriguing features for $B$ below 0.2 T. For
frequency higher than few GHz, the high-$B$ data exhibit behaviors
similar to a 2DES, showing IQH states. The SdH oscillations become
less pronounced at lower $f$ and even completely disappear below
600 MHz. In Fig.\ 2 (c), the data are shown up to $B=$ 11 T. An
extra adsorption peak appears at $B$$=$6 T associated with a
unique phase change nearby for $f$$=$255 MHz, and moves to lower
$B$ for higher $f$,  indicating that this feature relates to the
edge magnetoplasma\cite{Grodnensky1994} (EMP) excitations. More
detailed data and explanation of these interesting results will be
published separately. Here we will emphasize mainly the resolution
of the measurements and also the polarization of each QW that can
be extracted from our data.

The background noise of the data shown in Fig.\ 2 (b) and (c) is
extremely low. The amplitude and phase fluctuations in $\Delta A$
and $\Delta\phi_{s}$ data are less than 0.003\% and
0.001\textdegree\ for $f$$\lesssim$10 GHz, and 0.05\% and
0.03\textdegree\ for $f$$\gtrsim$10 GHz, respectively.  These
resolution limits actually depend on the power reaching the LNA,
which is $f$ dependent due to the loss of the sample and the
semirigid cables. The average power into the sample and into the
LNA are about -70 dBm and -76 dBm at $f$$\sim$1 GHz, and down to
-80 dBm and -107 dBm at 14 GHz, respectively. Moreover, the 0.3 \%
scale bar in the $\Delta A$ plot is equivalent to only 53 nS 2D
conductivity in average, which is very small compared to the
signal levels in previous
studies.\cite{Engel1993,Li1997,Ye2002,Lewis2002,Chen2003,Ye2002B}

 The susceptibility $\gamma$ of each QW and $\Delta\phi_{s}$ can be
related\cite{peqderive} by $ \gamma=\Delta \phi_{s}\xi^{2}/N\omega
Z_{0}, $ where $N$ is the total number of QW's, $Z_{0}$ the
characteristic impedance of the CPW, and $\xi$ a length scale
related to the distribution of the tangential electric field ($E$)
on the surface and the geometry of the CPW. The induced dipole
moment of each QW segment, $p$, is then $\gamma E$. For our CPW
structure, $\xi$ is about 21 $\mu$m. The $\gamma$ value
corresponding to the 0.2\textdegree\ scale bar in Fig.\ 2(c) for
255MHz is about 3$\times$$10^{-27}$ F/m$^2$. For a signal of
$-$51.5 dBm peak power, we can estimate $p$ accordingly to be
about 3$\times$$10^{-25}$ Cm, equivalent to about 17 electrons
being transferred across a 0.1 $\mu$m effective QW width, assuming
a 0.3 $\mu$m depletion length near each edge.

Finally, we want to discuss the effect of the cable length and
related instrumental considerations. Usually as we increase $L$,
the sensitivity in phase is increased according to Eq.\ 1, and so
is the loop gain of the PLL. However, if $L$ is too big, the PLL
will have a very small capture range, and the effect of noise and
drift in electronic components become serious. In addition,
high-frequency signals will suffer a very severe loss.

In conclusion, we have developed and demonstrated a
high-sensitivity vector detection system for very low-power
microwave signals used in a CPW broadband sensor. This system is a
very powerful tool in studying the dynamic behaviors, including
the electric polarizations, of LDES's at low temperatures.

This work was supported by National Science Council of Republic of
China.


\begin{thebibliography}{99}
\bibitem{peqderive} This equation can be derived simply by equating
the energy stored in the field for the electric dipole moment of
all the QW's, $N\gamma E^{2}/2$, and the energy increment due to
change of the effective capacitance per unit length, $\Delta
CV^{2}/2$. The length scale $\xi$ can be shown to be
$[2(C_{a}/\epsilon_{0})\langle V^{2}/E^{2}\rangle]^{1/2}$, where
$C_{a}$ is the capacitance of the CPW if the substrate is replaced
by air, and $\langle V^{2}/E^{2}\rangle$ is the average of square
of voltage signal over square of tangential field on the surface
across the gap of the CPW.
\end{thebibliography}
\end{document}